# Three-point correlation function in the quasilinear regime

Y.P. Jing[1,2] and G. Börner[1,3]

[1] Max-Planck-Institut für Astrophysik, Karl-Schwarzschild-Strasse 1, 85740 Garching, Germany
[2] jing@mpa-garching.mpg.de
[3] grb@mpa-garching.mpg.de

**Abstract.** Using the second-order Eulerian perturbation theory (SEPT), we study the three-point correlation function $\zeta$ in the quasilinear regime for the SCDM, LCDM and MDM models, with the interesting result that these three models have distinctive three-point correlation functions. We test this SEPT prediction using a large set of high-resolution N-body simulations. The N-body results show that the SEPT prediction for $\zeta$ is not accurate even in the quasilinear regime ($r \lesssim 10\,h^{-1}$Mpc), in contrast to previous N-body tests on the skewness. However, similar to the perturbation theory, our N-body results still predict a strong dependence of the three-point correlation on the triangle shape which is observable in the distribution of galaxies if the galaxies trace the distribution of the underlying mass

## 1. Introduction

It is well known that the high-order correlation functions are very important cosmological measures which contain information in addition to the most widely used measure – the two-point correlation function $\xi$ (Peebles 1980). Determination of the high-order correlation functions, however, requires much better observational data than those needed for the two-point correlation function. Thanks to large angular and redshift surveys of galaxies which have been available recently or will be available, the high-order correlation functions have attracted much more attention.

*Send offprint requests to*: G. Börner

According to the Second-order Eulerian Perturbation Theory (hereafter SEPT, Peebles 1980), the skewness $S_3(R)$, which is related to the three-point correlation function $\zeta$ through

$$S_3(R) = \frac{\bar{\zeta}(R)}{\bar{\xi}^2(R)},$$
$$\bar{\zeta}(R) = \frac{1}{V^3} \int_{\text{sphere } R} d\boldsymbol{r}_1\, d\boldsymbol{r}_2\, d\boldsymbol{r}_3 \zeta(r_{12}, r_{23}, r_{31}), \quad (1)$$
$$\bar{\xi}(R) = \frac{1}{V^2} \int_{\text{sphere } R} d\boldsymbol{r}_1\, d\boldsymbol{r}_2\, \xi(r_{12}),$$

depends only on the shape of the linear power spectrum $P(k)$ if the primordial density fluctuation is Gaussian (Fry 1984; Bouchet et al. 1992; Juszkiewicz et al. 1993; Catelan et al. 1995). The SEPT prediction has been shown to be in very good agreement with the results of N-body simulations in the quasilinear regime (Juszkiewicz et al. 1993; Luchin et al. 1994; Bernardeau 1994; Baugh et al. 1995; Colombi et al. 1996). It is also expected that the skewness is a statistic sensitive to a possible bias of the galaxy distribution (Fry & Gaztañaga 1993; Mo et al. 1996) and to a possible non-Gaussianity of the initial density fluctuation (Fry & Scherrer 1994). The statistical analysis of the APM galaxy catalogue (Gaztañaga 1994, 1995) has yielded a skewness for *galaxies* which is compatible with the theoretical prediction for the *mass* skewness, provided that the primordial fluctuation is Gaussian and the power spectrum $P(k)$ is the same as that measured for the APM galaxies (Gaztañaga & Freiman 1994). An important implication is then that there exists little bias between the distributions of the galaxies and of the underlying mass. Mo et al. (1996) have recently studied the skewness for two plausible bias models that identify either primordial density peaks or dark matter halos as 'galaxies'. They found that for a power spectrum with a similar shape to that observed in the APM survey, the skewness of these 'galaxies' agrees with the APM galaxy skewness only when the spatial bias between the 'galaxies' and the underlying mass is small.

The three-point correlation function contains much richer information than the skewness, since the latter is an integral of the former (Eq.1). Based on SEPT, Fry (1984) calculated the three-point correlation function for scale-free power spectra $P(k) \propto k^n$. He pointed out that the normalized three-point correlation function $Q$, which is defined as

$$Q(r_{12}, r_{23}, r_{31}) = \frac{\zeta(r_{12}, r_{23}, r_{31})}{\xi(r_{12})\xi(r_{23}) + \xi(r_{23})\xi(r_{31}) + \xi(r_{31})\xi(r_{12})}, \quad (2)$$

depends on the shape of the triangle constructed from the three points $\mathbf{r}_1$, $\mathbf{r}_2$ and $\mathbf{r}_3$. The shape dependence in turn depends on the index $n$ of the power spectrum (see also Fry 1994 and §2). If the SEPT prediction holds for $\zeta$ in the quasilinear regime as for $S_3$ and if galaxies trace mass as the previous studies on $S_3$ suggest, we would expect such a shape dependence in the three-point correlation function of galaxies. This shape dependence would certainly be important for determining the primordial power spectrum and for understanding the bias processes. Therefore, it is very important to study and to test with N-body simulations the SEPT predictions for the three-point correlation function.

In this paper, we will calculate the three-point correlation functions for *realistic* power spectra in SEPT. We consider the power spectra of two Cold Dark Matter (CDM) models and one Mixed Dark Matter (MDM) model. The two CDM spectra, which are specified by the parameter $\Gamma \equiv \Omega h$ (where $\Omega$ is the current density parameter and $h$ is the Hubble constant in unit of 100km s$^{-1}$Mpc$^{-1}$, Bardeen et al. 1986), have $\Gamma = 0.5$ and $\Gamma = 0.225$ respectively. The first CDM spectrum is well known as the Standard CDM (SCDM) power spectrum, and the second is usually regarded as a power spectrum for a low-density CDM universe (LCDM). The MDM model assumes an Einstein–de Sitter universe with one of species neutrino with density $\Omega_\nu = 0.3$, CDM density $\Omega_{\rm CDM} = 0.6$, baryon density $\Omega_{\rm B} = 0.1$ and the Hubble constant $h = 0.5$. Both LCDM and MDM spectra are known to be compatible with the large scale structures observed in the local Universe, therefore it is very important to find statistics to distinguish between them (Bahcall 1995, Efstathiou et al. 1992, Jing et al. 1994, Klypin et al. 1993). A very encouraging result from this calculation is that the dependence of $Q(r_{12}, r_{23}, r_{31})$ on the triangle shape is so sensitive to the shape of the power spectrum that the LCDM and MDM spectra can hopefully be discriminated by the three-point correlation function in the quasilinear regime.

Although the SEPT prediction for the skewness has been shown to be valid in the quasilinear regime by a number of authors, there is no *a priori* reason to believe that the SEPT prediction for the three-point correlation function is equally valid since the skewness is related to the three-point correlation function through an integral (Eq. 1). As shown in recent work by Jing et al. (1995), two quite different forms of the three-point correlation function can result in an indistinguishable skewness. Therefore, it is valuable to use N-body simulations to test the SEPT prediction for $\zeta$. In this paper, we will use a large set of P$^3$M N-body simulations to test the SEPT prediction for $\zeta$. The N-body test will not only tell us whether the SEPT results of $\zeta$ can be applied to real observations but will also show us, as a result of general interest, whether the SEPT prediction for $\zeta$ agrees with N-body simulations as accurately as the prediction for the skewness.

## 2. Prediction of the perturbation theory

In the second-order Eulerian perturbation theory, the bispectrum $B(\mathbf{k}_1, \mathbf{k}_2, \mathbf{k}_3)_{\sum k_i=0}$, which is the Fourier transform of the three-point correlation correlation function, is predicted to be (Fry 1984)

$$B(\mathbf{k}_1, \mathbf{k}_2, \mathbf{k}_3)_{\sum k_i=0} = \left[\frac{10}{7} + \frac{\mathbf{k}_1 \cdot \mathbf{k}_2}{k_1 k_2}\left(\frac{k_1}{k_2} + \frac{k_2}{k_1}\right) + \frac{4}{7}\left(\frac{\mathbf{k}_1 \cdot \mathbf{k}_2}{k_1 k_2}\right)^2\right] P(k_1)P(k_2) \quad (3)$$
$$+ (\text{cyc.}).$$

Strictly speaking, the above expression is derived only for the Einstein–de Sitter universe. However it has been shown that this expression is also a very accurate approximation for $0.1 \leq \Omega < 1$ (Bouchet et al. 1992, Catelan et al. 1995), so we will apply this equation to the low-density flat model with $\Omega = 0.3$ as well (§3).

The three-point correlation function is then

$$\zeta(\mathbf{r}_1, \mathbf{r}_2, \mathbf{r}_3) = \sum_{\mathbf{k}_1, \mathbf{k}_2, \mathbf{k}_3} B(\mathbf{k}_1, \mathbf{k}_2, \mathbf{k}_3) e^{-i(\mathbf{r}_1 \cdot \mathbf{k}_1 + \mathbf{r}_2 \cdot \mathbf{k}_2 + \mathbf{r}_3 \cdot \mathbf{k}_3)}. \quad (4)$$

After some tedious calculation, we can write $\zeta(\mathbf{r}_1, \mathbf{r}_2, \mathbf{r}_3)$ as

$$\zeta(\mathbf{r}_1, \mathbf{r}_2, \mathbf{r}_3) = \left\{ \frac{10}{7}\xi(r_{21})\xi(r_{31}) \right.$$
$$- \left[\eta_2(r_{21})\eta_0(r_{31}) + \eta_0(r_{21})\eta_2(r_{31})\right] \mathbf{r}_{21} \cdot \mathbf{r}_{31}$$
$$+ \frac{4}{7}\left[\epsilon_2(r_{21})\epsilon_2(r_{31})(\mathbf{r}_{21} \cdot \mathbf{r}_{31})^2 + \epsilon_2(r_{21})\eta_2(r_{31})r_{21}^2 \quad (5)\right.$$
$$\left.+ \eta_2(r_{21})\epsilon_2(r_{31})r_{31}^2 + 3\eta_2(r_{21})\eta_2(r_{31})\right] \right\}$$
$$+ (\text{cyc.}),$$

where $\mathbf{r}_{ij} \equiv \mathbf{r}_i - \mathbf{r}_j$, $\xi(r)$ is the two-point correlation function

$$\xi(r) = \frac{1}{2\pi^2} \int P(k) \frac{\sin kr}{kr} k^2 \, dk, \quad (6)$$

and $\eta_l(r)$ and $\epsilon_l(r)$ ($l$ is 0 or 2) are respectively

$$\eta_l(r) = \frac{1}{2\pi^2} \int \frac{P(k)}{k^l} \frac{kr \cos kr - \sin kr}{kr^3} k^2 \, dk \quad (7)$$

and

$$\epsilon_l(r) = \frac{1}{2\pi^2} \int \frac{P(k)}{k^l} \times \frac{3(\sin kr - kr \cos kr) - k^2 r^2 \sin kr}{kr^5} k^2 \, dk. \quad (8)$$

For a power-law power spectrum $P(k) \propto k^n$, from the above expressions, one can easily get the result of Fry (1984), i.e.

$$\zeta(\mathbf{r}_1, \mathbf{r}_2, \mathbf{r}_3) = \left\{ \frac{10}{7} + \frac{4}{7}\left[\left(\frac{\gamma}{\gamma-3}\right)^2 \frac{1}{r_{12}^2}\frac{1}{r_{31}^2}(\mathbf{r}_{21}\cdot\mathbf{r}_{31})^2 - \frac{(2\gamma-3)}{(\gamma-3)^2}\right] \right. \\ \left. + \frac{\gamma}{\gamma-3}\left[\frac{1}{r_{12}} + \frac{1}{r_{31}}\right]\mathbf{r}_{21}\cdot\mathbf{r}_{31} \right\} \xi(r_{21})\xi(r_{31}) \\ + (\text{cyc.}), \quad (9)$$

where $\gamma \equiv n+3$ is the slope of $\xi(r)$ [$\xi(r) \propto 1/r^\gamma$]. For a realistic power spectrum, to calculate the three-point correlation function, one has to integrate Eqs. (6-8) numerically as we do below.

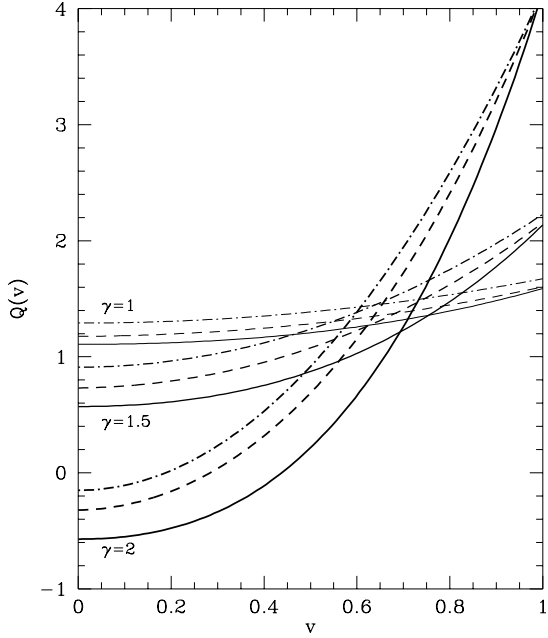

**Fig. 1.** The normalized three-point correlation function $Q$ as a function of $v$, predicted by the second-order perturbation theory for scale-free power spectra with $\gamma = 1$ (thin curves), 1.5 and 2 (thick curves). Solid lines are for $u = 1$, dashed lines for $u = 2$ and dot-dashed lines for $u = 10$.

The properties of $\zeta$ for scale-free power spectra are very instructive for understanding the results of $\zeta$ for realistic power spectra, therefore we first discuss $\zeta$ for scale-free power spectra. In this case, the normalized three-point correlation function $Q$ only depends on the shape of the triangle and on the index $n$ of the power spectrum. There are many ways to express the shape of a triangle. After some trials, we found that Peebles's variables $r$, $u$ and $v$ (Peebles 1980) provide a very good presentation of the three-point correlation function. For a triangle with three sides $r_{12} \leq r_{23} \leq r_{31}$, $r$, $u$, and $v$ are defined as:

$$r = r_{12}, \qquad u = \frac{r_{23}}{r_{12}}, \qquad v = \frac{r_{31}-r_{23}}{r_{12}}. \quad (10)$$

Clearly, $u$ and $v$ characterize the shape and $r$ the size for a triangle. With these variables, $Q$ does not depend on $r$ and depends very weakly on $u$ for a scale-free $P(k)$. The dependence of $Q$ on the triangle shape is then mainly the dependence on $v$. Figure 1 shows $Q$ as a function of $v$ for three scale-free spectra with $\gamma = 1$, 1.5 and 2 (or equivalently $n = -2, -1.5$ and $-1$). For each spectrum, $u$ is fixed to be 1, 2 or 10. The plot confirms a weak dependence of $Q$ on $u$. Depending on the index $n$, $Q$ may depend on $v$ strongly ($n = -1$) or weakly ($n = -2$), therefore the $v$-dependence provides a way to constrain the shape of the power spectrum.

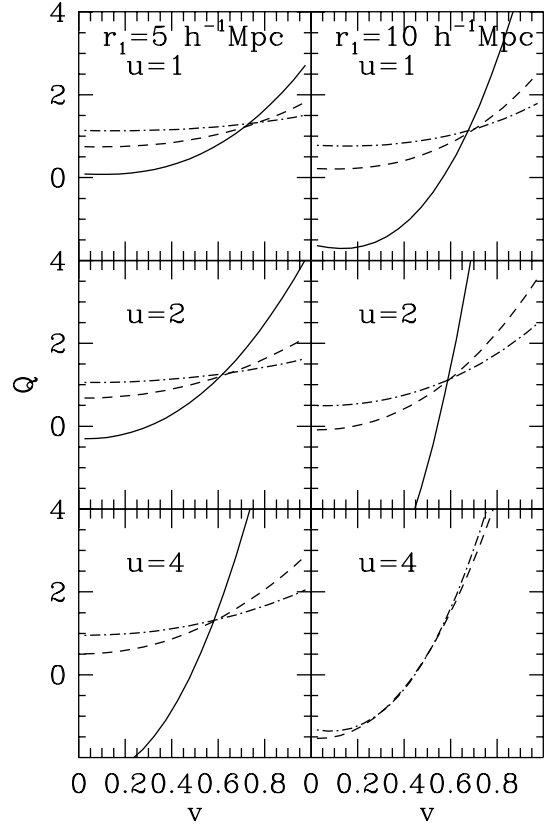

**Fig. 2.** The normalized three-point correlation function $Q$ as a function of $v$, predicted by the second-order perturbation theory for realistic power spectra. Solid lines are for SCDM, dashed lines for LCDM and dot-dashed lines for MDM.

In Fig. 2 we present our numerical results of $Q$ for three realistic power spectra. The SCDM ($\Gamma = 0.5$) and LCDM ($\Gamma = 0.225$) spectra are taken from Bardeen et al. (1986), and the MDM spectrum is the power spectrum given by Klypin et al. (1993) for cold dark matter at $z = 0$ in the MDM model of $\Omega_\nu = 0.3$. As in the case of scale-free power spectra, the normalized function $Q$ does not depend on the amplitude of the power spectrum. Here we are interested in the quasilinear regime (i.e. $r \gtrsim 5\,h^{-1}$Mpc) where the three-point correlation function can be estimated accurately both in observations and in N-body simulations and where the second-order approximation for $\zeta$ is likely to be valid according to previous simulation tests on $S_3$. In this regime, the free-streaming motion of neutrinos in the MDM model is not important and our treatment which considers only CDM is valid. The figure shows that the normalized three-point correlation function of SCDM is significantly different from those of LCDM and MDM. Even the LCDM and MDM models show sufficient difference in $Q$ for some configurations (e.g. $r = 10\,h^{-1}$Mpc and $u = 1$), so that one can hope to distinguish between these two interesting models through analyzing the three-point correlation function for a large galaxy survey. The results can be easily understood with the results for scale-free power spectra. For the scales $k$ we considered, the SCDM spectrum has the largest and the MDM the smallest effective index among the three spectra, therefore SCDM shows the strongest and MDM the weakest variation of $Q$ with $v$.

## 3. N-body simulation test

We use a large set of N-body simulations to test the SEPT results presented in the last section. The simulations were generated with a P$^3$M N-body code. A description of the code can be found in Jing & Fang (1994). Each simulation has $100^3$ to $128^3$ particles. To remove possible effects of a finite simulation volume, we used box sizes $L = 300$ or $400\ h^{-1}$Mpc. For each of the three power spectra, we have run 5 to 6 simulations in order to reduce the cosmic variances. Table 1 lists model and simulation parameters for these simulations.

**Table 1.** Parameters used in the simulations

|  | SCDM | LCDM | MDM |
|---|---|---|---|
| L ( $h^{-1}$Mpc) | 300 | 400 | 300 |
| No. of particles | $128^3$ | $100^3$ | $128^3$ |
| Force Resolution( $h^{-1}$Mpc) | 0.23 | 0.20 | 0.23 |
| No. of realizations | 5 | 6 | 5 |
| Initial redshift | 8 | 8 | 8 |
| Time steps | 400 | 400 | 400 |
| $\Omega_0$ | 1 | 0.3 | 1 |
| $\Lambda_0$ | 0 | 0.7 | 0 |
| Normalization $\sigma_8$ | 1.24 | 1 | 0.67 |

The three-point correlation functions for simulation particles are estimated by

$$\zeta(r,u,v) = \frac{DDD(r,u,v) - DDR(r,u,v)}{RRR(r,u,v)} + 2, \qquad (11)$$

where $DDD(r,u,v)$ is the count of triplets with shape $(r,u,v)$ in the simulation, $RRR(r,u,v)$ is the count of triplets expected if these particles were randomly distributed and $DDR(r,u,v)$ is the expected count of triplets formed by two simulation particles and one random point (Peebles 1980). In this paper we will use the procedure of Jing et al. (1995) to estimate $\zeta$. Briefly, we estimate the two-point correlation function for particles and fit it by an analytical formula. Then we calculate $DDR(r,u,v)$ and $RRR(r,u,v)$ *analytically*. With counting triplets, we get $\zeta(r,u,v)$ through Eq. (11).

It is non-trivial to count triplets for more than $10^6$ particles with modern computers. Here we use the linked-list technique (Hockney & Eastwood 1981) to search for triplets. We limit our analysis to triangles with the longest side $r_{31}$ less than 0.1 box side length (i.e. $r_{31\,\mathrm{max}} = 0.1L$) for several reasons. First and most importantly, the main aim of this paper is to test the SEPT prediction for $\zeta$ in the quasilinear regime, i.e. $r \gtrsim 5\,h^{-1}$Mpc. Second, on scales larger than $0.1L$, the three-point correlation function is very weak and it is not easy to measure it. Furthermore, the finite-volume effect becomes more and more important (see discussion at the end of this section). Finally the computational time grows rapidly with $r_{31\,\mathrm{max}}$ ($\propto r_{31\,\mathrm{max}}^6$). However, even with the linked-list technique and with this choice of $r_{31\,\mathrm{max}}$, it is still not feasible to count triplets for more than $10^6$ particles. Therefore we randomly select 100,000 particles (still a very big number) from each simulation for our analysis.

Our N-body results of the three-point correlation function are presented in Figure 3. For all three models, the normalized function $Q$ increases with the parameter $v$, qualitatively in agreement with the SEPT prediction. Quantitatively, the degree of the agreement between the N-body result and the SEPT prediction seems to depend on the power spectrum. For the LCDM model the agreement looks best, and for the SCDM model the agreement is worst. The N-body results of the LCDM model agree quite well with the SEPT prediction except at $v \approx 0.95$ where the N-body result is significantly higher. The disagreement between the N-body result and the SEPT prediction is statistically significant ($> 2\sigma$) for all three models. In the MDM model, the N-body $Q$ shows a stronger $v$-dependence than the SEPT prediction. On the contrary, the SCDM simulations show a weaker $v$-dependence than SEPT predicts. As a result, the difference in the $v$-dependence among the models is smaller in the N-body simulations than the prediction by SEPT. This point can be clearly seen in Fig. 3c where we compare the N-body results of the three models.

Numerical artifacts of simulations and non-linear effects of evolution can both weaken the agreement between the simulation results and the perturbative predictions. Among many possible numerical artifacts, *only* the finite volume effect could have significantly influenced our simulation results since the scales we are interested in are above $5\,h^{-1}$Mpc. The finite volume effect is caused by the fact that the simulation cannot include density fluctuations on scales above the simulation box size. We can quantitatively estimate this effect on our results by calculating two SEPT values of $Q$. One is calculated as in §2, and the other is calculated with the lower integral limits of Eqs.(6-8) set to $2\pi/L$ the fundamental wavenumber of the simulation. For the triangle configurations in Fig. 3 and for the three power spectra studied in this paper, the difference between these two SEPT values is small (either the absolute difference is less than 0.1 or/and the relative difference is less 10%). Therefore, the finite volume effect on our results is negligible, and the non-

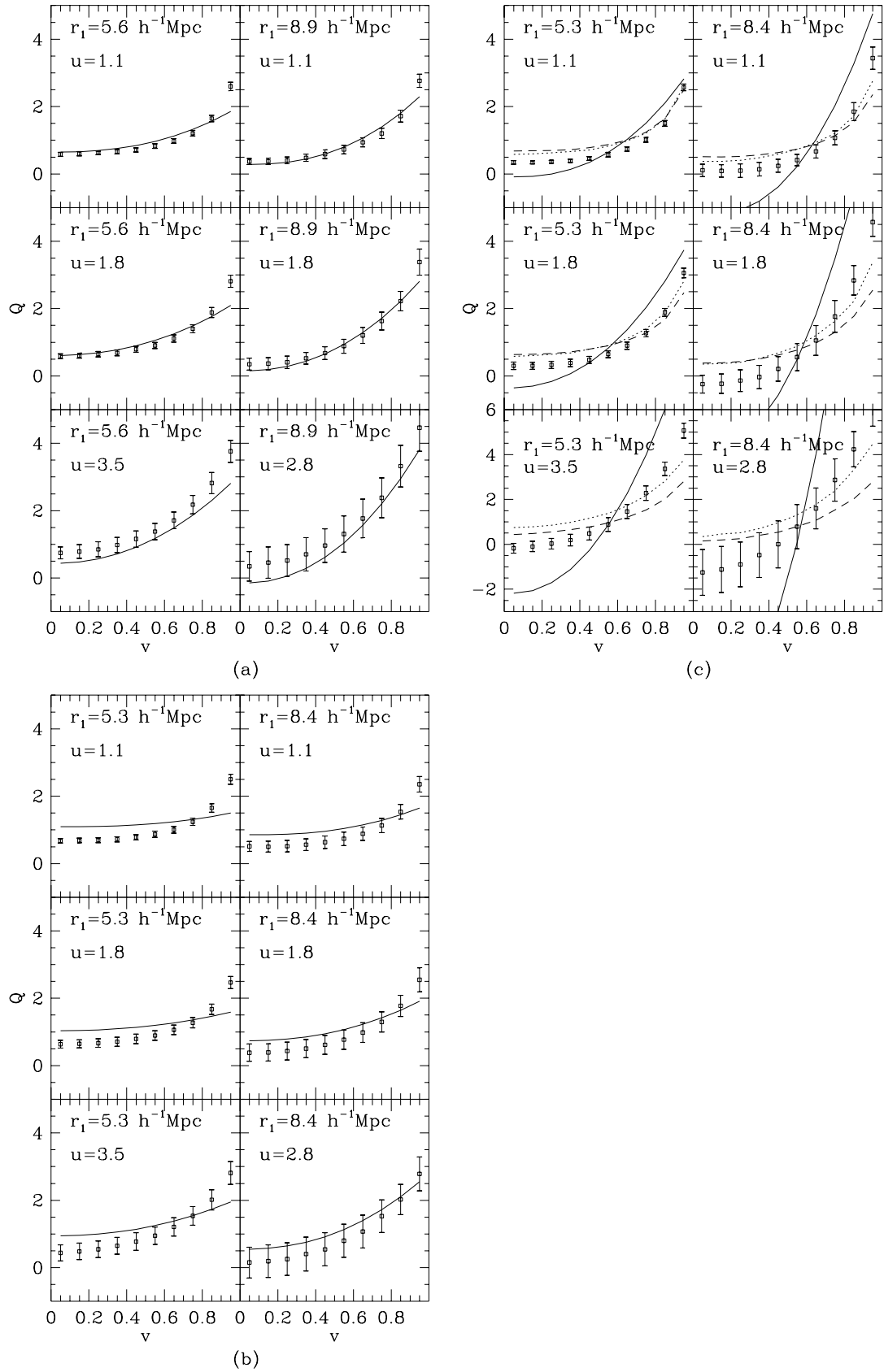

**Fig. 3.** The normalized three-point correlation function $Q(r, u, v)$ estimated from the N-body simulations (symbols), compared with the predictions by the second-order perturbation theory (the solid lines). The error bars are estimated from the fluctuation among different realizations. **a)** for LCDM; **b)** for MDM; and **c)** for SCDM. For comparison, we draw again the N-body results of LCDM (dotted lines) and MDM (dashed lines) on the SCDM plot.

## 4. Conclusion and further discussion

In this paper, we have carefully studied the three-point correlation function in the quasilinear regime both with the second-order Eulerian perturbation theory and with a large set of N-body simulations. With the perturbation theory, we showed that the normalized three-point correlation function $Q$ sensitively depends on the shape of the linear power spectrum. For scale-free power spectra with index $-3 < n < 0$, $Q(r, u, v)$ is an increasing function of $v$, and the larger $n$ (i.e. the flatter the spectrum), the more rapidly $Q$ increases with $v$. Because of differences among the effective slopes of the SCDM, LCDM and MDM spectra in the quasilinear regime, for a fixed configuration of a triangle, the SCDM model shows the strongest and the MDM the weakest variation of $Q$ with $v$. With these dependences, even the two popular models, LCDM and MDM, could perhaps be discriminated by measuring $Q$ for a large galaxy survey. Motivated by this potential importance, we analyzed the three-point correlation functions for a large set of N-body simulations. Our N-body results show that in the quasilinear regime ($r \lesssim 10 \, h^{-1}$Mpc), $Q$ increases with $v$ in all three models. Furthermore Q has the least increase with $v$ in the MDM model and has the most increase in the SCDM model. These two points qualitatively agree with the prediction of the perturbation theory. However, quantitatively, the N-body results show less variations of $Q$ with $v$ in the SCDM model and more variations in the other two models than the prediction of the perturbation theory. Thus the three-point correlation function is less powerful as a discriminator between the popular models than the perturbation theory originally suggests.

On the other hand, the robust dependence of $Q$ on $v$ predicted by our N-body simulations should exist in the distributions of galaxies if the galaxies really trace the underlying mass in spatial distribution. Search for this dependence will eventually put constraints on the bias. We note that Groth and Peebles (1977) have found a weak dependence of $q$ (the normalized three-point correlation function in angular distribution) on $v$ for the Lick catalogue. However, in their analysis, they have mixed (averaged) $q$ of large size and small size triangles. Since $q$ is approximately a constant for small triangles where clustering is strongly non-linear (Efstathiou et al. 1988, Matsubara & Suto 1994; we have confirmed this result in our simulations), their results cannot be directly compared with our N-body results here. We also note that a similar dependence of $Q$ on $v$ has been found for clusters of galaxies in N-body simulations (Jing et al. 1995) which means that the spatial distributions of clusters are closely related to the distribution of the underlying mass.

Besides many investigations which have tested with N-body simulations the high-order correlation prediction of the perturbation theory through the skewness (see §1), there are two publications which tested the bispectrum (Eq. 3) and the function $\zeta$ (Eq. 9) for scale-free power spectra. Fry et al. (1993) analysed the bispectrum for an ensemble of $128^3$ staggered-mesh simulations. Their results in the quasilinear regime agree with the SEPT prediction (Eq. 3), although the statistical errors in their N-body results are rather large (because of the small amount of independent triangles in the Fourier space in their analysis). More closely related to our work, Matsubara & Suto (1994) analyzed $\zeta$ for a set of Tree N-body simulations with $64^3$ to $128^3$ particles. They concluded that Eq.(9) disagrees with their N-body results, consistent with our results here. It is worth pointing out that the discrepancy they found looks more severe than the one we found here. The reason is probably that their comparison between theory and simulations is not optimal, since they did not separate out the non-linear regime where the SEPT prediction is expected to be invalid and they ignored the finite-volume effect which is serious for separations greater than 0.1 box size. Furthermore, their method for quantifying a triangle mixed configurations of different $v$ for a finite triangle-bin. For example, their triangles of shape-bin (1,10,10) could have all possible values 0 to 1 for $v$. Our discussions in §2 have shown that $Q$ depends very sensitively on $v$ in SEPT.

We have checked the skewness for our N-body simulations and found that the N-body skewness is in good agreement with the SEPT prediction, consistent with many earlier investigations (§1). Indeed, the skewness shows much better agreement between the perturbation theory and the N-body simulations than the three-point correlation function. The reason can be seen in our Fig. 3 where the N-body results of $\zeta$ cross the SEPT prediction curves at $v \approx 0.5$ for the two worse-fit models (SCDM and MDM). The skewness is an average of $\zeta(r, u, v)$ over $r$, $u$ and $v$ (Eq. 1) and the averaging improves the agreement for the skewness between the perturbation theory and the N-body simulations.

*Acknowledgements*. GB acknowledges support by SFB 375. YPJ would like to thank the Alexander-von-Humboldt Foundation for a Research Fellowship. He also thanks Max-Planck Institute für Astrophysik for financial support during the final stage of this work. We thank the referee, F.R. Bouchet, for insightful comments.